%% file: main.tex
\documentclass{article}
\usepackage[utf8]{inputenc}
\usepackage{graphicx}

\title{Towards a portable high-resolution muon detector based on Resistive Plate Chambers}

\author{S. Basnet$^1$ \and E. Cortina Gil$^1$ \and P. Demin$^1$ \and R.M.I.D. Gamage$^1$ \and A. Giammanco$^1$ \and R. Karnam$^1$ \and M. Moussawi$^1$ \and A. Samalan$^2$ \and M. Tytgat$^2$}

\date{$^1$Centre for Cosmology, Particle Physics and Phenomenology (CP3), Universit\'e catholique de Louvain,Louvain-la-Neuve, Belgium 
    \\ $^2$Department of Physics and Astronomy, Ghent University, Ghent, Belgium \\[3ex] \today}

\begin{document}

\maketitle

{\raggedleft CP3-22-07 \\
}

\begin{abstract}
\input{abstract}
\\
\\
{\it Proceedings of the International Workshop on Cosmic-Ray Muography (Muography2021), 24-26 Nov. 2022 at Ghent, Belgium. Submitted to Journal for Advanced Instrumentation in Science.}
\end{abstract}

\section{Introduction}
\input{intro}

\input{bulk_text}

\section*{Acknowledgements}
\input{acknowledgements}

\bibliographystyle{unsrt}
\bibliography{main}

\end{document}

%% file: abstract.tex
The use of conventional imaging techniques becomes problematic when faced with challenging logistics and confined environments. In particular, such scenarios are not unusual in the field of archaeological and mining explorations as well as for nuclear waste characterization. For these applications, even the use of muography is complicated since the detectors have to be deployed in difficult areas with limited room for instrumentation, e.g., narrow tunnels. To address this limitation, we have developed a portable muon detector (muoscope) based on glass Resistive Plate Chambers (RPC) with an active area of 16 $\times$ 16 cm$^{2}$. The specific design goals taken into consideration while developing our first prototype are portability, robustness, autonomy, versatility, safety and low cost. After gaining building and operating experience with the first prototype, We are currently in the process of developing an improved second prototype. In line with our design goals, we also study the possibility to switch the sensitive units from strips in the old prototype to pixels for the new one. This will help further improve our design goal of portability by reducing the overall weight of the setup by half since a single RPC layer provides bi-dimensional information with pixels. However, for performing high resolution muography, the number of readout units per layer will also need to increase significantly, leading to increase in the overall cost and power consumption of the muoscope. To mitigate these issues, we are developing a novel 2D multiplexing algorithm for reading out several pixels with a single electronic channel. In this article, we give an overview of the detector development, focusing mainly on the design goals and the choice of detector technology. Furthermore, we present the details of the expected changes in the new prototype as well as a simulated 2D multiplexing study based on general principles.

%% file: intro.tex
\label{sec:intro}

Cosmogenic muons have been used for the first time to probe the overburden of a tunnel in 1955~\cite{George1955}. 
Since then, muon tomography (also known as ``muography'') methods have found applications across many different fields~\cite{Bonechi:2019ckl}, including geosciences, archaeology, civil engineering, nuclear waste monitoring, etc. 
In some of these applications, it is not uncommon that the optimal place for installing the muon detector, from the point of view of imaging precision, is found in narrow and confined environments (e.g., tunnels, underground chambers and crevasses, crowded nuclear waste storage sites, etc.). Such places may occasionally come with additional logistic limitations, e.g. they may lack power supply. 

These factors have motivated a few muography teams to develop portable, compact and autonomous detectors. Examples include detectors based on scintillating bars~\cite{Baccani:2018nrn} or fibers~\cite{Kyushu2018}, which have the benefit of simplicity and reliability of operation, while being limited in the intrinsic resolution that they can achieve. Gaseous detectors, which can achieve high resolution relatively easily, are also in use~\cite{Lingacom2018,Schouten2018,roche2020muon} although they face a specific issue in this context because they are normally operated with a continuous gas flow~\cite{procureur2020we}, hence gas bottles need to be periodically replenished at the measurement site. Moreover, they may pose safety issues in confined environments due to the risk of anoxia, or to the presence of flammable or explosive components in their gas mixtures.

With these issues in mind, we aim at developing a portable muon telescope suitable for the imaging of small targets in contexts of challenging logistics. 
We chose glass Resistive Plate Chambers (RPCs) as detector units, upon considerations related to the good trade-off between resolution, efficiency, cost, and ease of construction
~\cite{MuographyBook}. 
We gained experience in building and operating a first prototype~\cite{Wuyckens2018,Basnet2020,Moussawi2021,Gamage:2021dqd}, that fulfill the requirements of portability, compactness and autonomy, and we lay the groundwork towards an improved second prototype.

This paper is organized as follows:
Section~\ref{sec:principles} exposes the guiding principles of our project, followed by a technical description of the current prototype in Section~\ref{sec:techdescription};
Section~\ref{sec:nxtprtyp} presents the ongoing developments, and 
Section~\ref{sec:mult} elaborates in particular on a dedicated multiplexing method.

%% file: bulk_text.tex
\section{Guiding principles}
\label{sec:principles}

This section elaborates on the main design goals, which are portability, versatility, safety and autonomy, and on the implications that they entail for a RPC-based detection setup. This section also summarizes the main technical choices of our current prototype; for more details, the reader is invited to read Refs.~\cite{Wuyckens2018,Basnet2020,Gamage:2021dqd}.

\begin{figure}[htbp]
\centering
\includegraphics[width=\textwidth]{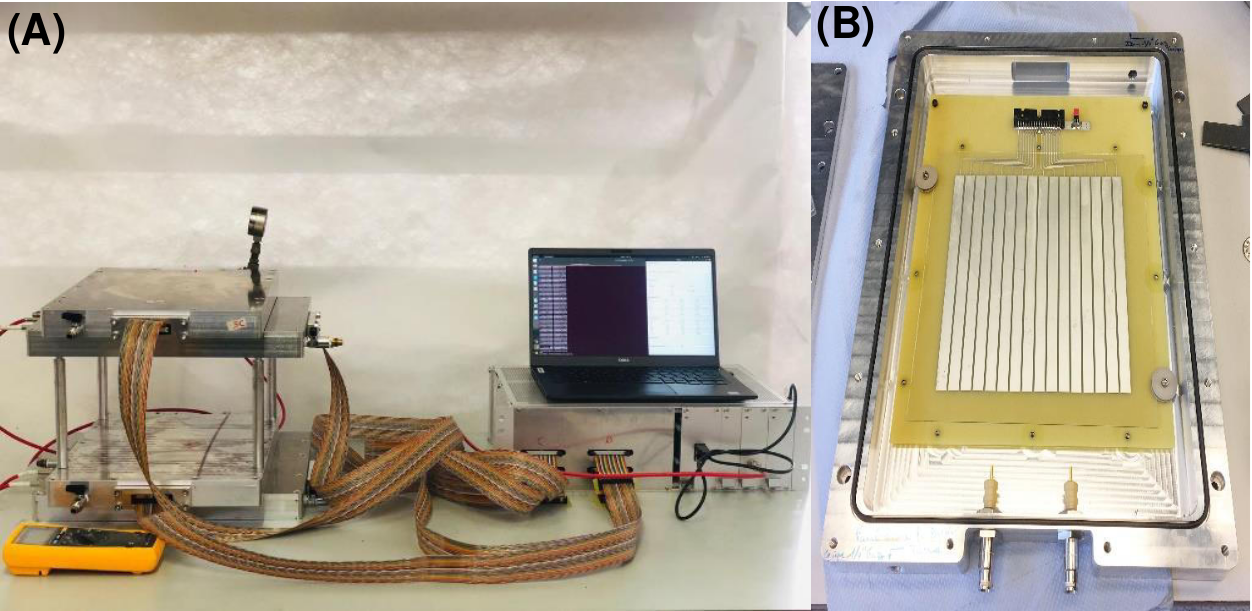}
\caption{\label{fig:detector} (A) Muoscope set-up consisting of four glass RPC layers and DAQ. (B) One of the RPCs inside its casing; it consists of 16 sensitive strips, hosted in an air-tight aluminum box.}
\end{figure}

Motivated by the requirement of portability, we are using small scale (active area of $16\times 16~cm^2$) RPCs. This small size is an unconventional choice for the standards of (astro)particle physics and muography, as typical motivations for using RPCs include, in fact, their relative ease of construction for large-area detectors and their relatively low cost per area~\cite{MuographyBook}. However, there are precedents for small-area RPCs in medical physics~\cite{Blanco2006}, although those are not intended to be portable. 
Portability also sets constraints on the weight of the complete setup. Currently, each RPC is hosted in a thick aluminum casing that weighs 6.5 kg. A complete standalone setup consisting of four identical RPCs is shown in Fig.~\ref{fig:detector} (A). The data acquisition system (DAQ) is integrated with the HV supply to the RPCs. 

Our project is not intended for a single type of application, therefore our setup must be versatile. In order to easily repurpose the setup for different types of measurements, we chose a modular design where the overall geometry is not fixed. 
Each RPC is housed in an individual aluminium casing and the different casings are separated by spacers, in arrangements that can be adapted to various use cases, as elaborated in Refs~\cite{Moussawi2021,Gamage:2021dqd}.
For example, four RPCs can be arranged in two adjacent pairs, alternating in orthogonal orientations, such that even and odd RPCs measure, respectively, the X and Y coordinates of the muon's passage. 
Ensuring safety in confined environments demands more constraints on gaseous detectors, which are typically operated with continuous gas flow using external gas supply (mainly through attached gas bottles); and autonomy means that the detectors should run for extended periods without human intervention, which means that gas refilling has to be as infrequent as possible (ideally only once, in the lab, before moving the setup to the place where it has to take data). 
The problem of reducing the necessary gas flow has been studied, for example, in Refs.~\cite{procureur2020we,Assis:2020mvp,nyitrai2021toward}. 
The aluminum casings that host our RPCs, shown open in Fig.~\ref{fig:detector} (b), are designed to be air-tight, allowing a stable operation during several weeks. The rate of gas leakage in vacuum conditions was measured using helium to be 10$^{-9}$ mbar~l~s$^{-1}$~\cite{Wuyckens2018}. 
Not needing gas bottles or gas flow allows us to use the muoscope in confined areas and also facilitates portability, by reducing weight and size of the overall setup.

\section{Current RPC prototype}
\label{sec:techdescription}

The gas mixture that constitutes the active detecting medium in our RPCs consists of R134a Freon (95.2\%), isobutane (4.5\%) and SF$_{6}$ (0.3\%), kept at a pressure slightly above (by $\sim$0.1 atm) the atmospheric one. 
The use of other mixtures is a possibility for the future, also taking into account that R-134a and SF$_6$ are environmentally unfriendly, and intense R\&D is being devoted in the RPC community to the search for new mixtures with a better trade-off between detector performance and global warming potential~\cite{Abbrescia:2016xdh,Guida_2020}. 

Glass sheets, 1~mm thick, are used as high-resistance parallel plates, and their exterior sides are painted with a semi-conductive coating that allows to spread high voltage (HV) throughout the plate and makes a uniform electric field across the gas volume. 
A uniform distance of 1.1~mm between the glass plates is obtained with nine round edge spacers made of polyether ether ketone (PEEK). 
The uniformity of the semi-conductive coating is important for the performances of RPCs, and in particular when they are used for muography~\cite{MuographyBook}. 
Therefore, a significant effort has been invested on this front. 
At the beginning of this project, we spread the paint manually with a paint-roller~\cite{Wuyckens2018}, a cheap procedure that has two major drawbacks: it does not scale well, and it can not ensure an excellent uniformity (we observed variations of up to 200\% in surface resistivity). 
Therefore, we produced a batch of glass plates where the paint was spread by serigraphy, whose variations in surface resistivity are now below 20\%~\cite{Basnet2020}. We monitor their surface resistivity regularly since almost two years, finding so far a slow drift in time but no variation in uniformity, and no visible correlation with environmental parameters~\cite{Gamage:2021dqd}.



RPC signals are picked up by 16 copper strips, 0.9~cm wide and separated by a 0.1~cm gap, meaning a pitch of 1~cm.
Data have been so far acquired via two front-end boards (FEBs) originally developed for the RPCs of the CMS experiment~\cite{FEB1, FEB2}, which can handle 32 analog inputs channels each. It is to be noted that at the start of the project the choice of only 16 strips for our four planes was dictated by the availability of only two FEBs, and not by the intrinsic position resolution (which is known to be potentially better by large factors for RPCs~\cite{Blanco2006}). 
Each channel consists of an amplifier with a charge sensitivity of 2 mV/fC, a discriminator, a monostable and a LVDS driver. The LVDS outputs of all the FEBs are connected to a System-on-Chip (SoC) module, which is installed on a carrier board with a wireless connection, to ensure autonomy also from the point of view of data transfer. 

\section{Towards Improved and High Resolution Next Prototype}
\label{sec:nxtprtyp}

While the purpose of this first prototype was just to gain building as well as operating experiences, the next prototype will have to pave the way towards our main aim, i.e., performing high resolution muography. Reachable goals for RPCs are $O(1mm)$ for intrinsic spatial resolution and $O(1ns)$ for timing~\cite{MuographyBook}. In order to achieve these resolution goals, a significant increase in a number of sensitive units, and thus, electronic channels, will be required. 

\begin{figure}[htpb]
\centering
\includegraphics[width=0.75\textwidth]{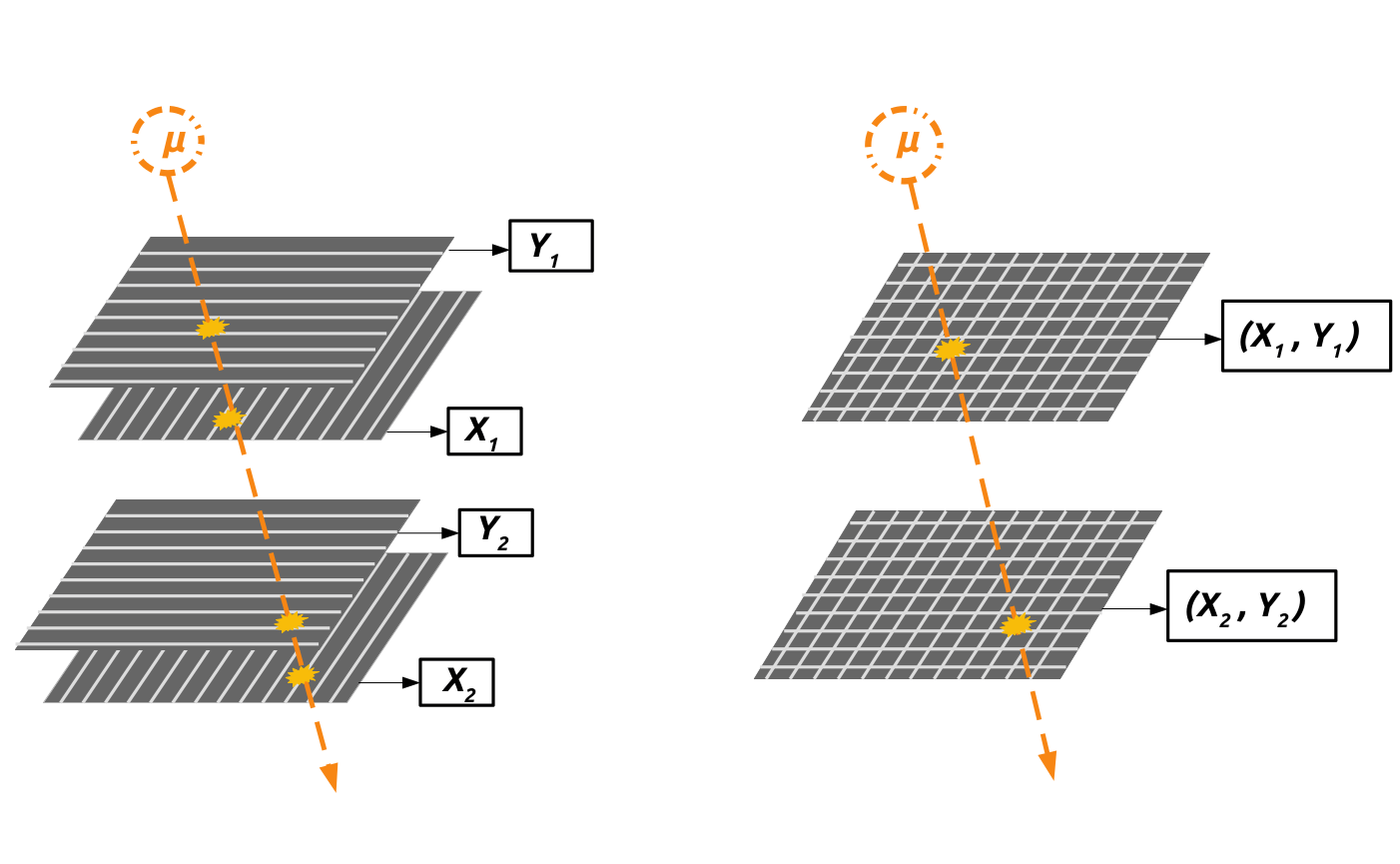}
\caption{\label{fig:multiplex}An illustrative setup with 4 layers of strips (left) and one with 2 layers of pixels (right) providing the same x, y information at the same two positions in z.}
\end{figure}

On the hardware side of things, we are planning to switch from our current CMS chip to MAROC chip~\cite{Barrillon:1091460}. The size of the MAROC chip is much smaller compared to CMS chip (O(1cm) vs O(10cm), respectively), which bodes well with our overall design goal of easy portability, and most importantly, the MAROC chip is also capable of handling eight times as much electronic channels: 8 channels per CMS chip in comparison with 64 channels per MAROC chip. Furthermore, the new chip will enable to exploit the timing information of the muon hits in our analysis of the muoscope data and we will have, for our next prototype, an opportunity to use timing information for better muon track selection as well as for background rejection. We already have access to the MAROC chip and are currently in the testing phase with the development kit with an aim to design a board dedicated and optimized for our muoscope.

Another important consideration being made for the new prototype is a possible switch from strips as sensitive units in our current version to pixels in the next one (see Figure \ref{fig:multiplex}). The obvious advantage of this switch, keeping in mind the design goals discussed earlier, is that both the number of layers and the total weight of the overall setup will be halved. Additionally, the total detector efficiency also will also improve; $\epsilon^{n/2} > \epsilon^{n}$, where $\epsilon$ is the efficiency of a single detector layer. On the other hand, the total power consumption as well as the cost of electronics could be negatively impacted because of the switch. The total cost and power consumption of the electronics scales linearly for the strips option whereas, for the pixels option, they scale quadratically. For the same spatial resolution, the total number of electronic channels required for the muoscope is 4$\times$N for the strips compared to 2$\times$N$^2$ for the pixels, where N is the segmentation along X and Y directions (hence, the number of strips for a 1D detector). For our current resolution and size, this means increasing the total number of electronic channels from 64 (strips) to 512 (pixels). 

It is clear that the only major counter-argument for opting the pixels solution is the overall cost of the electronics and one way to address this issue would be to read out several sensitive units (pixels, in this case) to a single electronic channel. This method of reading multiple units with one electronic channel is known as multiplexing~\cite{procureur2013genetic} and in order to make the cost of the strips and pixels option equal, one has to multiplex 10 pixels with one electronic channel. A brief feasibility study of a 2D (i.e., pixel) multiplexing scheme with a worked-out example is discussed in the following section.

\section{Multiplexing}
\label{sec:mult}

In the context of our muoscope, multiplexing is to be done with a combination of hardware (i.e., how the pixels are connected) and software (i.e., optimal clustering of adjacent pixel signals). Even though multiplexing depends equally on both of these aspects, we are currently studying how the pixels should be connected while a thorough study of an optimal de-multiplexing technique will be performed in the near future. 

Our current approach is to randomly generate possible multiplexing mapping schemes that fulfill relatively simple rules for the pixel’s connection~\cite{ourPatentApp}. The only rule we impose in our procedure is that the sets of multiplexed pixels that are connected together are evenly distributed in the plane and cannot be adjacent (including the edges). Further optimization of the hardware connections is to be carried out studying multiple mapping schemes and determining the one with least spurious clusters generated after multiplexing.

\begin{figure}[h!]
\centering
\includegraphics[width=0.75\textwidth]{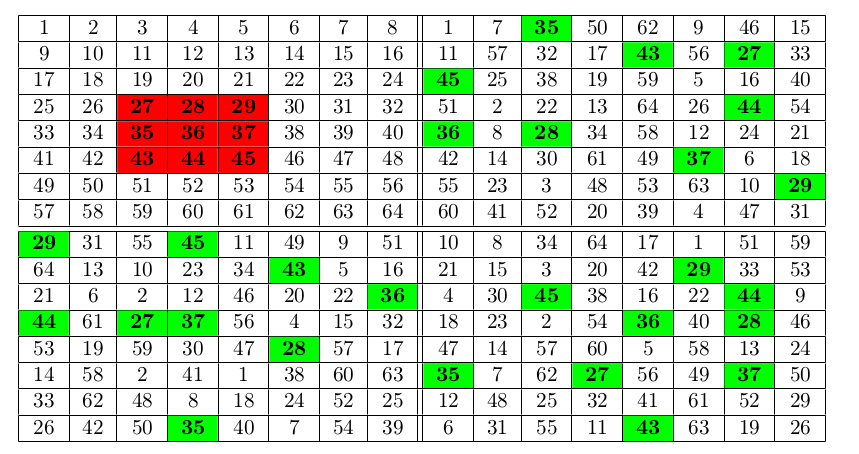}
\caption{\label{fig:gridmap} An example multiplexing scheme where a bigger grid of 16$\times$16 pixels is sub-divided, with double lines, into 4 grids each with dimensions 8$\times$8 pixels (i.e., multiplexing factor of 4). The numbers in the matrix represent the pixel labels. The pixels with the same labels from these 4 sub-matrices are to be connected together.}
\end{figure}

Generally, a few main parameters defines multiplexing. For our case, these parameters are listed below.
\begin{itemize}
    \item Total number of pixels in the matrix ($N_{x}\times N_{y} = 16\times 16$);
    \item Expected Cluster Size ($C_{x} \times C_{y} = 3\times3$);
    \item Multiplexing factor ($M$): number of pixels connected to each other, hence number of sub-grids;
    \item Total number of readout channels (N$_{r}$).
\end{itemize} 

However, not all of these parameters are independent of each other. In particular, the total number of readout channels is simply the ratio of the total number of pixels in the matrix over the multiplexing factor.

As an example on how the pixel connection part of the multiplexing process would look like, firstly we studied an easy case with a matrix of size 16 $\times$ 16 pixels (256 pixels in total) and a relatively mild multiplexing factor, $M=4$, which means 64 readout channels ($N_{r}$) in total. We follow a "divide and conquer" approach where we divide our main 16 $\times$ 16 matrix into 4 sub-matrices (each with 8 $\times$ 8 pixels). Following the rule of adjacency from above, we created a multiplexing mapping scheme such that the pixels from all four sub-matrices with the same labels are going to be connected together. A representative example of such a mapping scheme for $M=4$ is shown in Figure \ref{fig:gridmap}. Since the arguments of cost and power consumption of the electronics demand $M\ge 8$ for making pixels more convenient than strips, we have, in addition to $M=4$, studied the case where $M=8$, i.e. where the main grid is divided into 8 sub-matrices, each with 4 $\times$ 8 pixels and $N_{r}=32$.

\begin{figure}[htbp]
\centering
\includegraphics[width=\textwidth]{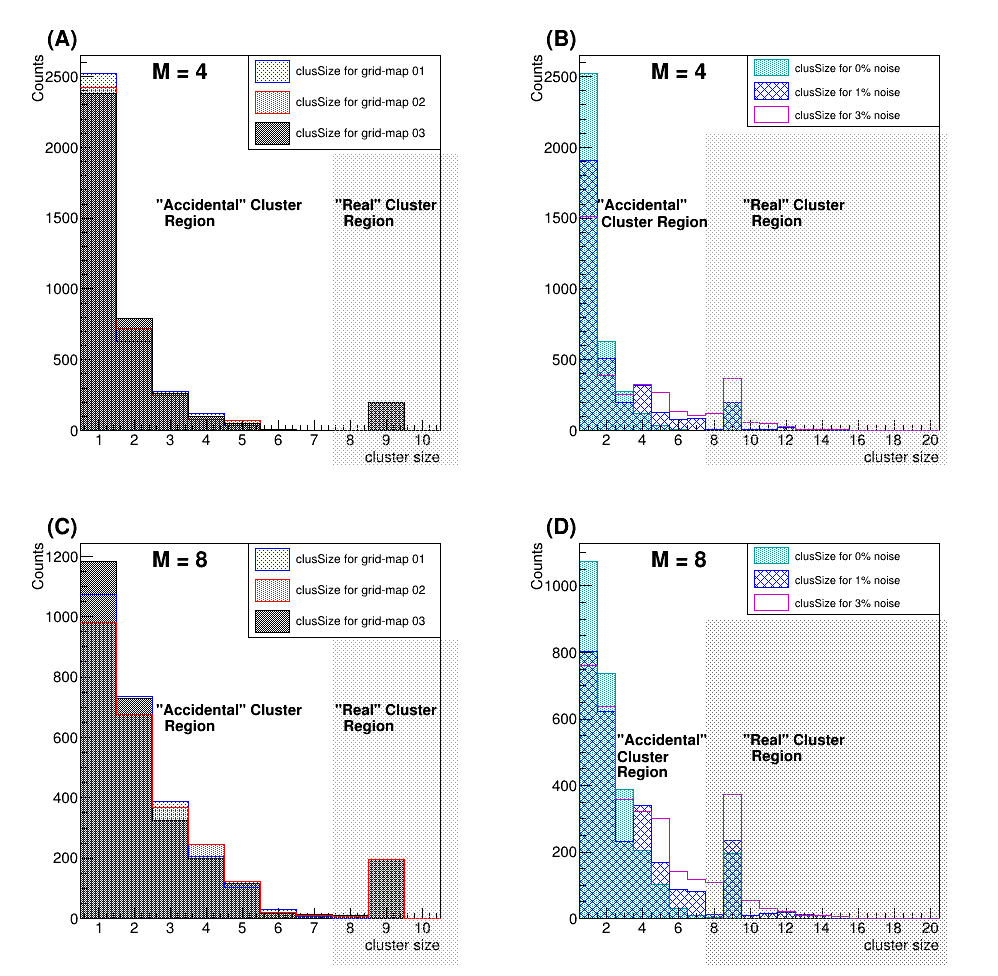}
\caption{\label{fig:multiplex2} Cluster size distributions comparing three different randomly-generated grid maps as well as various levels of noise for a single grid map (no noise, 1\% and 3\%). Top panels (A and B) and bottom panels (C and D) show results for multiplexing factors $M=4$ and $M=8$, respectively.}
\end{figure}

For de-multiplexing, we exploit the spill-over of the "real" signal over neighboring channels (as in e.g. Ref.~\cite{procureur2013genetic}). Our adjacency rule is meant to ensure that a set of adjacent pixels in the true hit location can not correspond directly to another set of adjacent pixels in other sub-grids. However, accidental clusters of adjacent pixels may still be formed, especially with the help of stochastic noise. 
It should also be noted that we do not take into account cases with multiple muons per event; however, these are a rare occurrence when the detector area is small as is our case.
We expect to be able to tune the geometrical parameters, and therefore the signal spill-over, such that a typical "real" signal to spread over immediately adjacent pixels, therefore a typical cluster to be made of 3$\times$3 pixels. All possible single muon signals (14$\times$14 = 196) were generated in the main grid and the resulting cluster size data was produced. In Figure \ref{fig:gridmap}, an example 3$\times$3 signal in the first sub-matrix is shown in the red cells and the corresponding pixels with the same labels in the rest of the sub-matrices are highlighted with green cells. For this signal, we have 1 "accidental" cluster of size 3, 1 "accidental" cluster of size 2, and 22 "accidental" size 1 clusters as well as 1 "real" cluster of size 9. The cluster size for all possible 196 signals for three different grid maps are compared for $M$ of 4 and 8 are shown in Figures \ref{fig:multiplex2}(A) and \ref{fig:multiplex2}(C), respectively.

In order to examine the effects of stochastic noise to our multiplexing approach, we have also simulated, along with 3 $\times$ 3 "real" signals, 1\% and 3\% (which is very pessimistic) (i.e., 8 extra pixels, ) noisy pixels in this study, which for this grid size means roughly 3 and 8 extra noisy pixels, respectively, on average. 
The resultant cluster size distributions for a specific grid map ("grid-map 01"), comparing 0\%, 1\% and 3\% noise cases for $M$ of both 4 and 8, are shown in Figures \ref{fig:multiplex2}(B) and \ref{fig:multiplex2}(D), respectively. 

From Figure \ref{fig:multiplex2}(A) and (C), it can be clearly seen that the "accidental" clusters rapidly die out as the cluster size increases, dividing the cluster size distributions into two distinct regions: "real" and "accidental" cluster regions (as labelled in the figure). Although a comparison of only three distinct grid maps is shown in this current study, the aforementioned trend seems to be more or less universal as we could observe it in all the O(100) grid maps that we generated for this study. 
However, increasing levels of stochastic noise would eventually lead to a situation where the "real" and "accidental" cluster regions begin to be indistinguishable. 
To find out what level of noise can be tolerated by our method, we show the cluster size distributions corresponding to 3\% (1\%) noise in blank magenta (shaded teal) histograms in Figures \ref{fig:multiplex2}(B) and \ref{fig:multiplex2}(D). The blurring  between "real" and "accidental" regions due to noise is more pronounced for $M=8$ than for $M=4$, as qualitatively expected. In comparison with the simpler no noise cases where no events with cluster size greater than 9 was seen, the tail of the cluster size distributions after the addition of stochastic noise is relatively longer, in some cases extending to the size of 16. These features in the distributions clearly exhibit that a simple cluster size based discrimination is not robust enough for de-multiplexing offline and begins to fail with the introduction of stochastic noise. Therefore, as a future development, we intend to make our de-multiplexing procedure exploit the expectations about the cluster shape; this will demand a detailed simulation of the signal formation and of the cross-talk between channels. 

Another important parameter that has not not been explored in this study, but might be worth studying, is the ratio of total grid size and the $M$ factor. Moreover, our current study uses an arbitrary "real" signal size. More grounded results require a realistic estimate on the "real" signal size as well as level of noise expected with pixels, which can only be possible after the fabrication of a pixel based readout board that is already in the pipeline. A more thorough and systematic study with more aggressive multiplexing (factor of $> 8$), taking into account also the cluster-shape, will follow soon.


\section{Conclusion \& Outlook}

We reported on the current status of our project for the development of a portable, compact and versatile muon detection system for muography. 
Our technological choices take into account the possibility that such a setup needs be operated in a confined space, possibly with challenging logistics.

Our system, based on mini glass-RPC detectors, is intended to be low cost and portable, not only in terms of size and weight but also with respect to gas tightness and ease of transportation of the full setup, including electronics. 
After gaining experience with our first prototype, and having addressed the problem of the uniformity of the resistive coating by the usage of serigraphy, we are laying the groundwork towards a second generation detector with state-of-the-art electronics. For our next prototype, we aim at better spatial resolution and at the introduction of 2D reading (i.e., switching from strips to pixels). 
To avoid that these new developments lead to an explosion in the cost and power consumption of the electronics, we are developing a dedicated method for 2D multiplexing.

%% file: acknowledgements.tex
This work was partially supported by the EU Horizon 2020 Research and Innovation Programme under the Marie Sklodowska-Curie Grant Agreement No. 822185, and by the Fonds de la Recherche Scientifique - FNRS under Grant No. T.0099.19. The corresponding author also acknowledges additional research grants from the FNRS - FRIA. We thank Dr. Stephan Aune and his serigraphy team at CEA, Saclay for helping us with resitive coating. We also thank the electronics group at the Center for Cosmology, Particle Physics, and Phenomenology (CP3), Universit\'{e} catholique de Louvain, for their help.